\pdfoutput=1

\documentclass[]{spie}

\usepackage{amsmath,amsfonts,amssymb}
\usepackage{graphicx}
\usepackage{siunitx}
\usepackage[colorlinks=true, allcolors=blue]{hyperref}
\usepackage{multirow,hhline}

\title{Upgrading the Submillimeter Array: wSMA and beyond}

\author[a]{Paul K. Grimes}
\author[a]{Garrett K. Keating}
\author[a]{Raymond Blundell}
\author[a]{Robert D. Christensen}
\author[a]{Mark Gurwell}
\author[a]{Attila Kovacs}
\author[a]{Timothy Norton}
\author[a]{Scott N. Paine}
\author[a]{Ramprasad Rao}
\author[a]{Edward C.-Y. Tong}
\author[a]{Jonathan Weintroub}
\author[a]{David Wilner}
\author[a]{Robert W. Wilson}
\author[a]{Lingzhen Zeng}
\author[a]{Qizhou Zhang}
\affil[a]{Center for Astrophysics \textbar\ Harvard \& Smithsonian, 60 Garden St, Cambridge, MA, USA}

\authorinfo{Further author information: (Send correspondence to P.K.G.)\\P.K.G.: E-mail: pgrimes@cfa.harvard.edu, Telephone: 1 617 496 7640\\  G.K.K.: E-mail:garrett.keating@cfa.harvard.edu, Telephone: + 1 617 495 7487}

\begin{document} 
\maketitle

\begin{abstract}
The Submillimeter Array (SMA) is an array of 8 antennas operating at millimeter and submillimeter wavelengths on Maunakea, Hawaii, operated by the Smithsonian Astrophysical Observatory and Academia Sinica Institute of Astronomy and Astrophysics, Taiwan. Over the past several years, we have been preparing a major upgrade to the SMA that will replace the aging original receiver cryostats and receiver cartridges with all new cryostats and new 230 and 345 GHz receiver designs. This wideband upgrade (wSMA) will also include significantly increased instantaneous bandwidth, improved sensitivity, and greater capabilities for dual frequency observations. In this paper, we will describe the wSMA receiver upgrade and status, as well as the future upgrades that will be enabled by the deployment of the wSMA receivers.
\end{abstract}


\section{INTRODUCTION}
\label{sec:intro}  % \label{} allows reference to this section
The Submillimeter Array (SMA) is an array of 8 antennas operating at millimeter and submillimeter wavelengths on Maunakea, Hawaii, operated by the Smithsonian Astrophysical Observatory and Academia Sinica Institute of Astronomy and Astrophysics (ASIAA), Taiwan\cite{Ho2004}. First commissioned in 2003, the SMA has been operating for over 20 years and has been continually upgraded with new receiver elements, polarimetric capabilities, enhanced frequency coverage, and a steadily increasing correlator bandwidth. At present, the frequency coverage of the SMA is from 180 to 420 GHz (\qtyrange{1.67}{0.710}{\unit{\mm}}). 

The SMA is a highly productive, in-demand astronomical facility, with a 3:1 oversubscription rate for open-skies observing, and has enabled over 1000 publications with 50,000 citations. Core areas of science observations include star formation, protoplanetary disks, nearby and high-redshift galaxies, galaxy clusters, evolved stars, the intersteller medium, super-massive blackholes and solar system objects. The flexibility of the SMA operations has enabled a significant contribution to time-domain astronomy, both for monitoring and for target-of-opportunity observations. The SMA is also a key station in the Event Horizon Telescope\cite{Akiyama_2019}.

Over the past several years, we have been preparing a major upgrade to the SMA that will replace the aging original receiver cryostats and receiver cartridges with all new cryostats and upgraded dual-polarized 230 GHz and 345 GHz receivers\cite{Grimes2020_wSMA_receivers}. This wideband SMA (wSMA) upgrade will also include significantly increased instantaneous bandwidth, improved sensitivity, and greater capabilities for dual frequency observations\cite{Wilner2017}. The deployment of the physically smaller wSMA receiver system will also create space in the SMA receiver cabins for additional optics and receivers, allowing the deployment of additional bands and guest instrumentation. The first prototype wSMA receiver was recently installed into an SMA antenna for on-sky testing and integration into the array, while the first production receivers are being readied for deployment to the SMA site.

In this paper, we will describe the wSMA receiver upgrade and status, as well as the future upgrades that will be enabled by the deployment of the wSMA receivers. In particular, we envision that the dual band 230/345 GHz wSMA receiver will be augmented with additional simultaneous frequency bands, starting with an upcoming 100 GHz receiver band, and later extending to higher frequencies to build the Panchromatic SMA\cite{Keating2023}.

\section{CURRENT STATUS OF THE SMA}
\label{sec:current}
\begin{figure}
    \centering
    \includegraphics[width=0.5\textwidth]{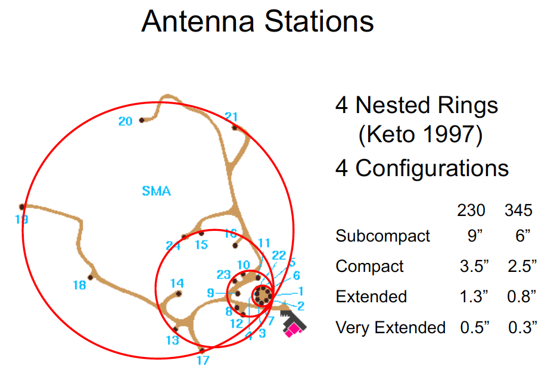}\includegraphics[width=0.5\textwidth]{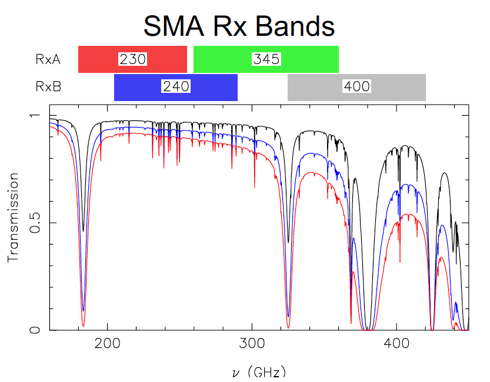}
    \caption{(left) The SMA consists of eight 6m antennas that can be arranged in four configurations with baselines of up to 508m, each with different angular resolutions. (right) Frequency coverage of the current four single polarized receivers, overlaid on 75\%, 50\% and 25\% percentile atmospheric transmission from Maunakea. The overlap in frequency coverage between the 230 and 240, and 345 and 400, receiver bands allows for polarimetric observations.}
    \label{fig:sma_current}
\end{figure}

The Submillimeter Array (SMA) is an 8-element radio interferometer located atop Maunakea in Hawaii at 4080m AMSL. Operating at frequencies from 180 GHz to 418 GHz, the 6m dishes may be arranged into several configurations, producing a range of spatial resolutions (Fig.~\ref{fig:sma_current}). The SMA operates in 4 configurations: "Subcompact", with maximum baselines of 30m for the six antennas on the inner ring; "Compact" with baselines up to 70m; "Extended" with baselines up to 220m; and "Very Extended with baselines up to 508m. Each antenna can observe with two receivers simultaneously, with up to 12 GHz bandwidth each per sideband, for a total of 48 GHz on-sky instantaneous bandwidth. The SWARM digital correlator backend provides a uniform resolution of 140 kHz\cite{Primiani2016} for all observations, which is archived by the CfA's Radio Telescope Data Center at full spectral resolution for future archival research.

Each antenna houses a single cryostat with four single-polarized double-sideband (DSB) receivers. Two receivers are assigned to each of the two polarizations and the two IF signal paths to the correlator, labelled RxA and RxB, and selected by room temperature receiver selection optics in front of the cryostat. All eight antennas are currently equipped with four receivers each, named 230GHz(RxA), 240GHz(RxB), 345GHz(RxA), and 400GHz(RxB). The on-sky tuning ranges for these receivers are shown in Table \ref{table:sens}. Simultaneous dual frequency operation is possible by pairing one RxA and one RxB receiver. Thus, the current full complement of receivers allows dual frequency operation for the receivers 230/240, 230/400, 345/240 and 345/400 (but not 230/345 or 240/400). 

The 230 and 240 GHz receivers overlap in LO coverage from 210 GHz to 240 GHz, the 240 and 345 GHz receivers from 258 GHz to 270 GHz, and the 345 and 400 GHz receivers from 336 GHz to 351 GHz, and thus can be used to observe both polarizations at the same frequency at the same time. For the 230 and 240 GHz, and 345 and 400 GHz receiver pairs, quarter-wave plates can be introduced into the optics and full-Stokes polarimetery carried out for both spectral lines and continuum observations\cite{Marrone2008}.

\section{THE wSMA UPGRADE}
\label{sec:wsma}
\begin{figure}
    \centering
    \includegraphics[width=\textwidth]{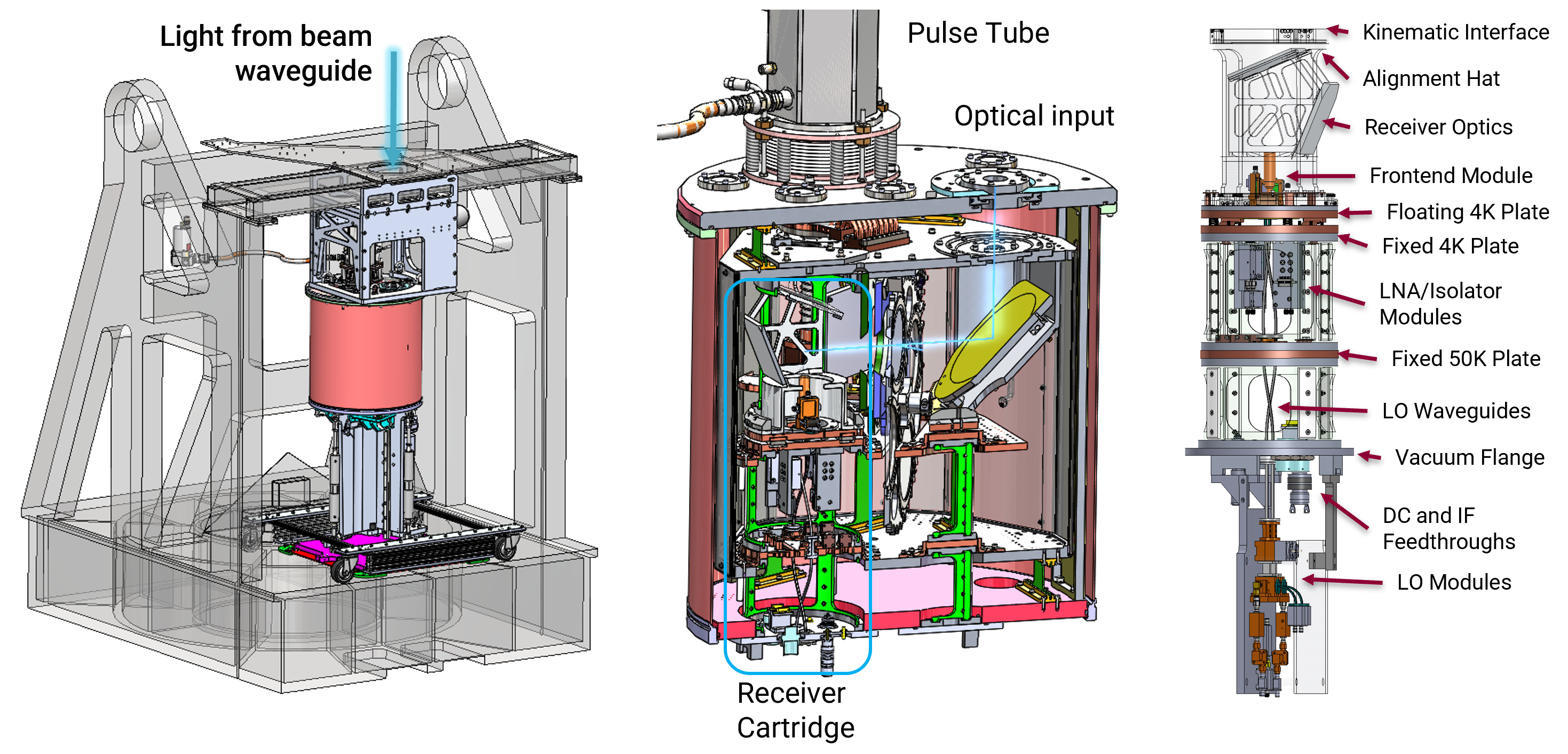}
    \caption{(left) CAD model of the wSMA receiver cryostat in the SMA antenna cabin. (center) Cut-through of the wSMA cryostat, showing the optical path to the receiver cartridges (light blue) (right) wSMA receiver cartridge.}
    \label{fig:wsma_CAD}
\end{figure}

The initial concept of the wSMA upgrade was developed in 2015-16, when it became clear that the end of the service lifetime of the receiver cryostat compressors and cryocoolers was approaching. The wSMA upgrade consists of the replacement of the existing SMA 8-cartridge (4 cartridges in use) cryostat, receiver cartridges and receiver optics with a new two-cartridge cryostat (Fig.~\ref{fig:wsma_CAD}) incorporating cooled receiver optics, upgrades to the local oscillator and receiver control electronics, and upgrades to the ambient calibration load/quarter-waveplate assemblies. At the same time, upgrades to the realtime control system are being implemented with the introduction of the SMA-X \cite{Kovacs2024} system. 

The wSMA receiver cryostats contain the the receiver selection optics, cooled to $\sim$\qty{50}{\kelvin}, and two receiver cartridges, with the final receiver focusing optics being mounted on the floating 4 K plate of the two cartridges.  Each receiver cartridge is inserted from the bottom of the cryostat, and cooled via two automatic thermal links, one in the lower radiation shield, and one in the 4 K plate. The upper radiation shield plate of the cryostat is used as an optics bench for the receiver optics, with the input fold mirror, receiver selection mechanism and cold calibration load mounted to plate. The receiver front-end and receiver cartridge optics are automatically aligned on insertion to the cryostat by a G-10 alignment ``hat'' that mates the floating 4 K stage of the cartridge to three ball and vee-groove kinematic mounts on the upper radiation shield plate. The cryostats are cooled by a Cryomech PT407-RM or PT410-RM pulse tube cryocooler, driven by the Cryomech CP286i variable frequency compressor. The variable frequency compressor allows the cooling power and power dissipation of the cryostat to be modulated, allowing power savings and limiting the requirements on cooling water in the antennas, while also providing for rapid cooldown.

The two receiver cartridges cover the two main frequency bands requested for observations with the SMA - 194-286 GHz and 264-376 GHz. Each cartridge houses a dual-polarized receiver front-end consisting of a profiled corrugated feed, ortho-mode transducer (OMT), LO couplers and two double-sideband SIS mixers\cite{Zeng2024}. Up to four IF output chains can be installed in each cartridge, allowing future upgrades to the use of diplexed IF chains that extend the IF band down to ~0.1 MHz\cite{Blundell2019}, or the installation of sideband separating receivers\cite{Garrett2023}. Each dual-polarized receiver cartridge replaces two of the current single polarized receiver cartridges (approximately SMA-230 and SMA-240 for the wSMA-Low receiver cartridge, and SMA-345 and SMA-400 for the wSMA-High receiver cartridge). The use of dual-polarized feeds and OMTs will greatly improve the optical alignment between orthogonal linear polarizations, and thus simplify polarimetric observations with the SMA.

\begin{figure}
    \centering
    \includegraphics[width=0.6\textwidth]{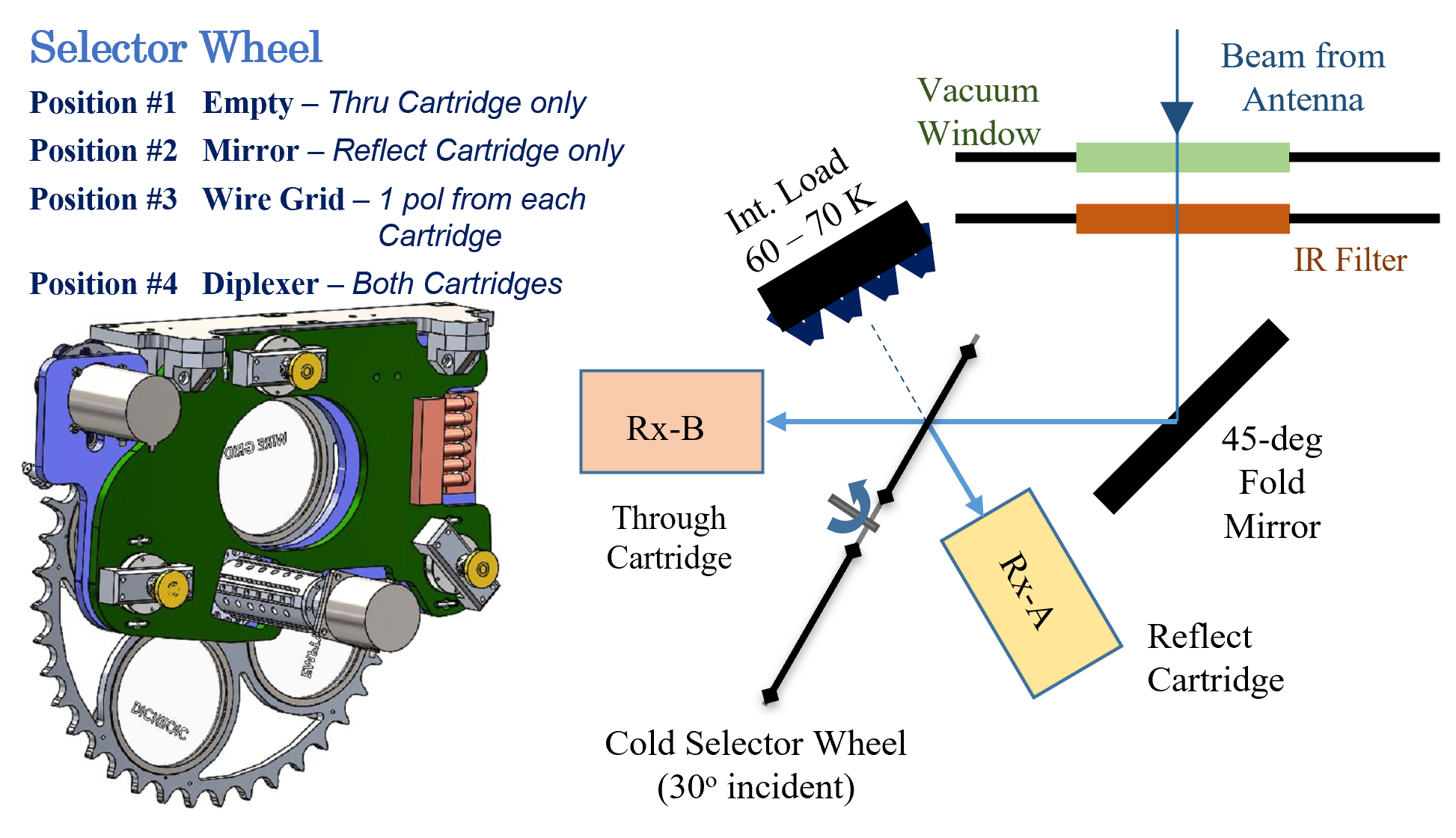}
    \caption{(left) CAD model of the wSMA receiver cryostat selector wheel mechanism that selects which receiver is coupled to the sky. (right) Schematic of the optical layout of the wSMA receiver. The receiver channels not coupled to the sky are coupled to an internal cold load, again mounted to the upper radiation shield plate of the cryostat.}
    \label{fig:wsma_selector}
\end{figure}

\begin{figure}
    \centering
    \includegraphics[height=3.5in]{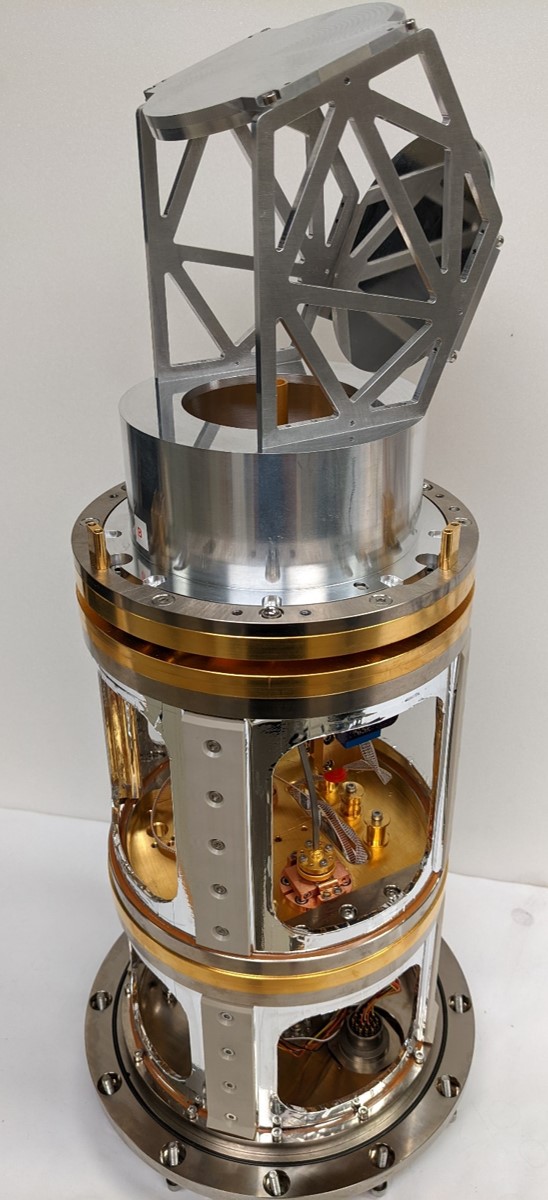}\hfill\includegraphics[height=3.5in]{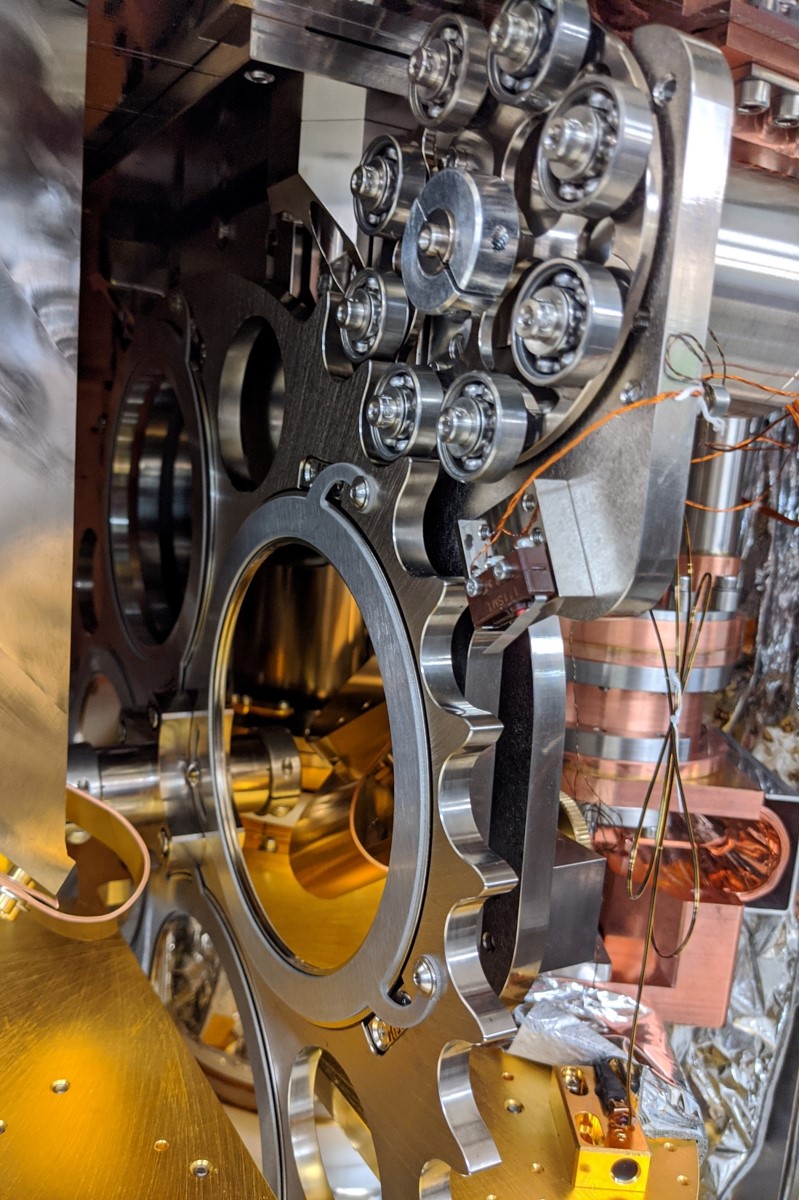}\hfill\includegraphics[height=3.5in]{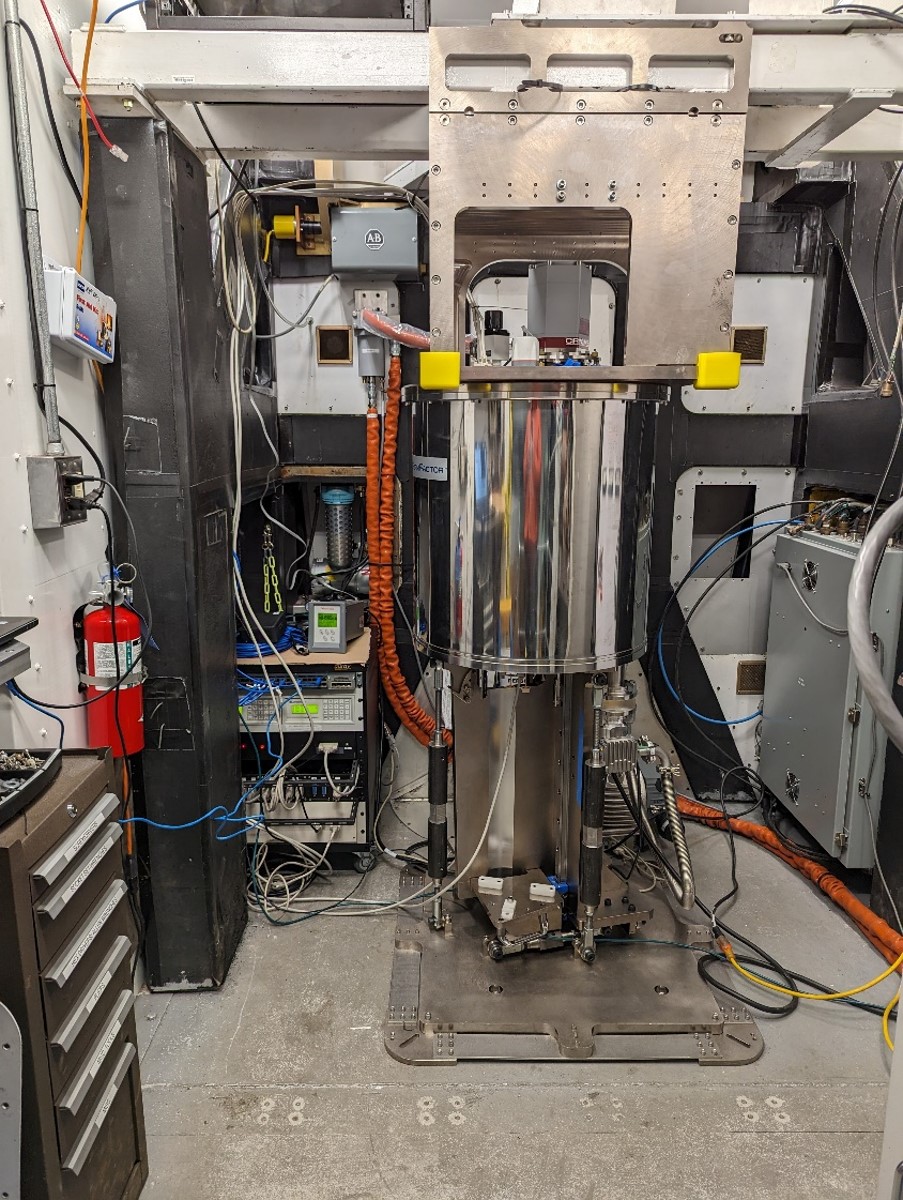}
    \caption{(left) wSMA Prototype receiver cartridge, populated with front-end receiver module, LO waveguides, and IF cables. (center) Selector wheel drive mechanism and selector wheel. (right) wSMA Prototype receiver installed in SMA Antenna 7.}
    \label{fig:wsma_hardware}
\end{figure}

The receiver that will see the sky is selected by a "selector-wheel" mechanism inside the cryostat radiation shield, mounted to the upper radiation shield plate, and cooled to $\sim$\qty{50}{\kelvin} (Fig.~\ref{fig:wsma_selector}). The selector wheel can hold up to four optical elements, nominally consisting of: a) an open aperture that allows the signal to travel to the ``through'' receiver cartridge; b) a plane mirror that redirects the incoming signal to the ``reflect'' cartridge; c) a wire polarizing grid, that sends one polarization of the signal to each receiver cartridge; and d) a dichroic beamsplitter, that sends different wavelengths of both polarizations to the two cartridges. The filter wheel is mounted on long axle with cryogenic rated silicon nitride bearings, while a "clock spring" arrangement of copper foil heat straps thermally connect it to the 1st stage of the cryostat. It is moved by a gear wheel holding eight cryogenic silicon nitride dry bearings that engage into shaped teeth on the filter wheel circumference, and driven by a cryogenic stepper motor (Fig.~\ref{fig:wsma_hardware}, center).

Over the past four years, two prototype cryostats have been tested in the Submillimeter Receiver Lab in Cambridge, and in the SMA Receiver Lab in the SMA's summit control building on Maunakea. During this time, we have developed the cryogenics, control systems, and most vitally, the receiver front-ends and cartridges (Fig. \ref{fig:wsma_hardware}, left). Earlier this year, the first prototype receiver system was installed into SMA Antenna 7, which was out of operation due to elevation drive failure, for extended field testing\cite{Grimes2024}. After a two week installation process, the prototype receiver has been operating near continuously in the field. On-sky testing and incorporation into SMA array observations will begin once the elevation drive repair is completed.

The lab testing and deployment of the prototype cryostats identified several design issues that needed to be rectified. The most important of these were: a) there was insufficient thermal margin in the design; and b) it was hard to achieve adequate optical alignment with only common-mode adjustment of the filter wheel, due to thermal distortion of the wheel on cooling to \qty{50}{\kelvin}. For the production cryostats, these issues have been resolved by replacing the PT407-RM cryocooler with a PT410-RM cryocooler, driven by the same CP286i inverter helium compressor; and by replacing the bulk adjustment of the filter wheel angle and position with individual adjustment of each optical element in the wheel. 

After implementation of these changes, two upgraded production cryostats were delivered to the Submillimeter Receiver Lab in December 2023. These cryostats are undergoing commissioning and optical alignment before shipping to Maunakea for installation in SMA antennas\cite{Christensen2024}. Two further production cryostats have been ordered, with an option for two further cryostats. The construction and deployment of the final two of the eight required cryostats is expected in 2026/27.

\section{THE PANCHROMATIC SMA}
\label{sec:panchromatic}
Building upon the baseline plan for wSMA (hereafter wSMA Stage I) described above, we have proposed a wSMA Stage II upgrade. Stage II is focused on adding a separate 3 mm LNA based receiver system alongside the wSMA Stage I receivers, and implementing tri-band observing modes. By adding room temperature and cryogenic dichroic beamsplitters\cite{Carter2024} to the SMA beam waveguide and to the wSMA Stage I cryostat's selector wheel, and providing additional IF paths, wSMA Stage II will be capable of simultaneous dual-polarization wideband observations at 3 \unit{\mm}, 1.3 \unit{\mm}, and 850 \unit{\um}. 

The addition of in-antenna IF digitization, IF transmission and processing capability will also allow us to implement sideband-separating receivers\cite{Garrett2023} using digital IF hybrids in the 1.3 \unit{\mm} and 850 \unit{\um} bands to improve sensitivity, particularly in less than optimum weather conditions (up to 4mm PWV, which is at 75th percentile of weather performance for Maunakea), and allow for fully simultaneous operation of all three receiver bands. With a total on-sky spectral grasp in excess of 100 GHz (0-16 GHz IF for 3 mm, 0-20 GHz for 1.3 mm and 850 $\mu$m; two sidebands and full-polarization), and with single frequency spectral line mapping speed improved by a factor of five over the current SMA performance, both continuum mapping and spectral line survey speeds will be improved by factors significantly greater than that.

Looking further into the future, we have proposed a wSMA Stage III concept we are calling the Panchromatic SMA\cite{Keating2023}, that aims for near-contiguous spectral coverage across the atmospheric windows from 3 \unit{\mm} up to 450 \unit{\um}, at high spectral resolution. This development would be powered by novel technologies currently under development at SAO, such as RF dichroic beamsplitters and diplexers, Tensor core GPU-based digital backends, and others which leverage the world-class expertise of SAO for receiver engineering and digital signal processing.

Stage III of the wSMA upgrade would utilize technological developments in wide band submillimeter technology and in reflectionless diplexers \cite{Morgan2015}, which are paving the way for a new class of receivers which provide broad simultaneous frequency coverage of the sky over a wide spectral range. In such a setup, the incoming beam from the telescope is first channelized into octave frequency bands through the use of dichroic beamsplitters. The beam for each of the octave bands is then captured by a wide-band smooth-wall feed horn, which will then feed a reflectionless filter bank, the output of which could be coupled to independent SIS mixers. We also anticipate that low noise amplifier technology will become dominant up to the 200 GHz range in the near future, enabling the use of cryogenic low noise amplifier operating with a 20 K cryocooler for these lower bands of frequencies. 

Such technologies would enable the expansion to a hex-band observing system, which will support simultaneous observations at 3 mm, 2 mm, 1.3 mm, 1.1 mm, \qty{850}{\um}, and \qty{750}{\um}. Further support for observations at \qty{600}{\um} and \qty{450}{\um} could be provided through additional instrumentation installed in the guest port position of the wSMA cryostat. The total spectral grasp would approach a terahertz in total bandwidth, requiring the construction of a 1 THz correlator-beamformer. Such a system would capitalize on advances in commercially available hardware, particularly Tensor-core GPUs, as well as the software and digital signal processing expertise at the CfA, to build a truly scalable software correlator architecture.

These proposed developments will follow a staged approach that allows for continued SMA operations during upgrades, and that continuously introduces additional capabilities for users of the SMA. An approximate ordering of the proposed upgrades is as follows, although many of the upgrades are independent of those before and after them and can be reordered or dropped as priorities and resource availability develops over time.

\subsection*{wSMA Stage II}
\begin{itemize}
\setlength\itemsep{-.3em}
    \item Simultaneous dual-polarization operation of the wSMA Stage I receivers using the cold dichroic beamsplitter in the wSMA cryostat.
    \item In-antenna IF digitization to provide additional and scalable digital IF transport to the correlator.
    \item Addition of the 3mm LNA receiver, coupling optics (including dichroic beamsplitters) and additional correlator capacity to allow tri-band observing.
    \item Replace the DSB Stage I receivers with sideband separation receivers.
    \item Expansion of IF bandwidths down to close to DC.
\end{itemize}

\subsection*{wSMA Stage III}
\begin{itemize}
\setlength\itemsep{-.3em}    
    \item Expansion of the 3mm LNA receiver to dual band 3mm/2mm operation.
    \item Addition of high frequency \qty{600}{\um} and/or \qty{450}{\um} receiver in "Guest" receiver position.
    \item Introduction of dual band \qty{1.3}{\mm}/\qty{1.1}{\mm} receivers.
    \item Introduction of dual band \qty{850}{\um}/\qty{750}{\um} receivers.
\end{itemize}

\begin{figure}[!t]
    \centering
    \includegraphics[width=\textwidth]{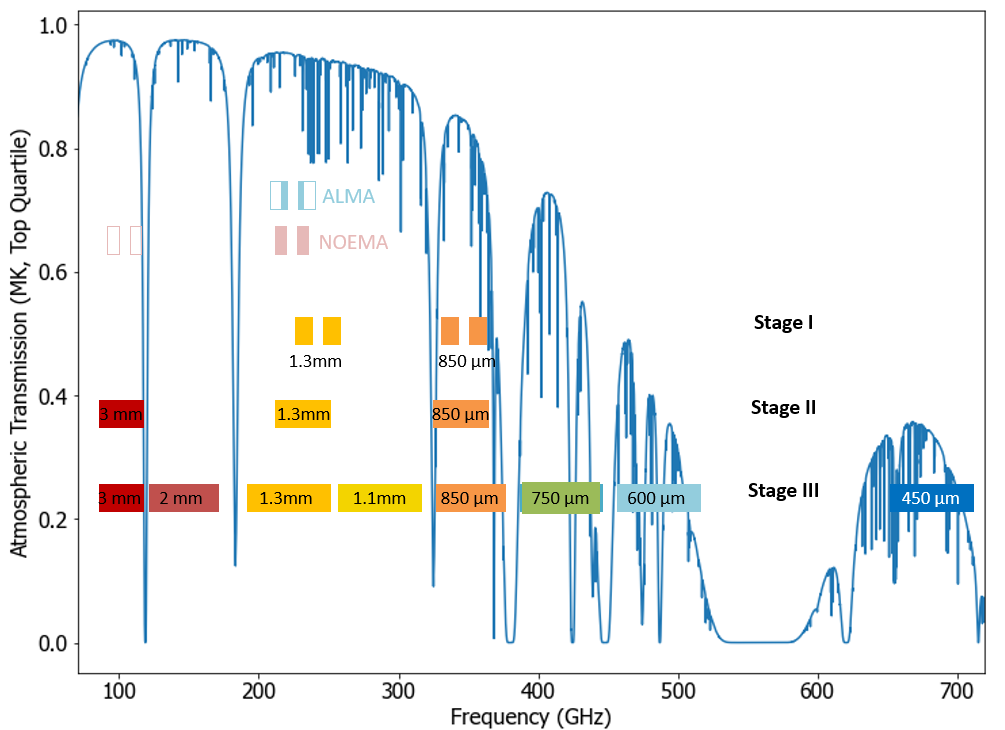}
    \caption{Instantaneous spectral coverage for example tunings of the various stages of the wSMA, along with some existing facilities. Compared with ALMA (currently 8 GHz instantaneous bandwidth, 32 GHz proposed for ALMA2030) and NOEMA (16 GHz instantaneous bandwidth, dual band observing proposed), Stage II and III would offer approximately one and two orders of magnitude greater instantaneous spectral coverage, respectively. Stage I of the wSMA upgrade has the same instantaneous bandwidth as the current SMA receivers.}
    \label{fig:spec_coverage}
\end{figure}

\begin{table}[!t]
\begin{center}
\begin{tabular}{|l|c|r|c|c|c|c|c|}\hline
            & Receiver     & \multicolumn{1}{c|}{RF}      & Inst. & Typ $T_{\rm sys}$ & Continuum  & \multicolumn{2}{c|}{Line Sensitivity} \\
  Milestone & Band         & \multicolumn{1}{c|}{Range}   & BW    & (SSB)             & Sensitivity  & \multicolumn{2}{c|}{1 km/s channel} \\
            &              & \multicolumn{1}{c|}{[GHz]}   & [GHz] & [K]               & [mJy/beam] & [mJy/beam] & [Jy km/s] \\ 

  \hhline{========}
  \multirow{4}{*}{Current SMA}
            & ``230''      & 172-254 & 24    & \phantom{0}222\phantom{$^{\textrm{c}}$}    & 0.221$^{\textrm{a}}$ & \phantom{0}39.0$^{\textrm{a}}$ & 0.0390$^{\textrm{a}}$ \\ \cline{2-8}
            & ``240''      & 198-278 & 24    & \phantom{0}222\phantom{$^{\textrm{c}}$}    & 0.221$^{\textrm{a}}$ & \phantom{0}39.0$^{\textrm{a}}$ & 0.0390$^{\textrm{a}}$ \\ \cline{2-8}
            & ``345''      & 242-367 & 24    & \phantom{0}702\phantom{$^{\textrm{c}}$}    & 0.755$^{\textrm{a}}$ & 104.1$^{\textrm{a}}$ & 0.1041$^{\textrm{a}}$ \\ \cline{2-8}
            & ``400''      & 320-424 & 24    & 1494\phantom{$^{\textrm{c}}$}              & 1.582$^{\textrm{a}}$ & 186.1$^{\textrm{a}}$ & 0.1861$^{\textrm{a}}$ \\ 
  \hhline{========}
  \multirow{2}{*}{Stage I}
            & 1.3 mm       & 194-286 & 24    & \phantom{0}147\phantom{$^{\textrm{c}}$}    & 0.128$^{\textrm{a}}$ & \phantom{0}22.5$^{\textrm{a}}$ & 0.0225$^{\textrm{a}}$ \\ \cline{2-8}
            & 850 \unit{\um}   & 264-376 & 24    & \phantom{0}484\phantom{$^{\textrm{c}}$}    & 0.405$^{\textrm{a}}$ & \phantom{0}55.7$^{\textrm{a}}$ & 0.0557$^{\textrm{a}}$ \\ 
  \hhline{========}
  \multirow{2}{*}{Stage II}
            & 3 mm         &  84-116 & 30    & \phantom{00}44\phantom{$^{\textrm{c}}$}    & 0.026\phantom{$^{\textrm{a}}$}      & \phantom{00}7.4\phantom{$^{\textrm{a}}$} & 0.0074\phantom{$^{\textrm{a}}$} \\ \cline{2-8}
            & 1.3 mm       & 190-290 & 40    & \phantom{0}107\phantom{$^{\textrm{c}}$}    & 0.052\phantom{$^{\textrm{a}}$}      & \phantom{0}12.0\phantom{$^{\textrm{a}}$} & 0.0120\phantom{$^{\textrm{a}}$} \\ \cline{2-8}
            & 850 \unit{\um}   & 320-380 & 40    & \phantom{0}358\phantom{$^{\textrm{c}}$}    & 0.167\phantom{$^{\textrm{a}}$}      & \phantom{0}29.5\phantom{$^{\textrm{a}}$} & 0.0295\phantom{$^{\textrm{a}}$} \\ 
  \hhline{========}
  \multirow{8}{*}{Stage III} 
            & 3 mm         &  84-116 & 32    & \phantom{00}44\phantom{$^{\textrm{c}}$}    & 0.026\phantom{$^{\textrm{a}}$} & \phantom{00}7.4\phantom{$^{\textrm{a}}$} & 0.0074\phantom{$^{\textrm{a}}$} \\ \cline{3-8}
            & + 2 mm       & 120-170 & 50    & \phantom{00}65\phantom{$^{\textrm{c}}$}    & 0.029\phantom{$^{\textrm{a}}$} & \phantom{00}9.0\phantom{$^{\textrm{a}}$} & 0.0090\phantom{$^{\textrm{a}}$} \\ \cline{2-8}
            & 1.3 mm       & 190-250 & 60    & \phantom{0}111\phantom{$^{\textrm{c}}$}    & 0.045\phantom{$^{\textrm{a}}$} & \phantom{0}12.5\phantom{$^{\textrm{a}}$} & 0.0125\phantom{$^{\textrm{a}}$} \\ \cline{3-8}
            & + 1.1 mm     & 255-315 & 60    & \phantom{0}162\phantom{$^{\textrm{c}}$}    & 0.063\phantom{$^{\textrm{a}}$} & \phantom{0}15.8\phantom{$^{\textrm{a}}$} & 0.0158\phantom{$^{\textrm{a}}$} \\ \cline{2-8}
            & 850 \unit{\um}   & 325-375 & 50    & \phantom{0}407\phantom{$^{\textrm{c}}$}    & 0.165\phantom{$^{\textrm{a}}$} & \phantom{0}32.9\phantom{$^{\textrm{a}}$} & 0.0329\phantom{$^{\textrm{a}}$} \\ \cline{3-8}
            & + 750 \unit{\um} & 385-445 & 60    & 1681\phantom{$^{\textrm{c}}$}              & 0.351\phantom{$^{\textrm{a}}$} & \phantom{0}97.2\phantom{$^{\textrm{a}}$} & 0.0972\phantom{$^{\textrm{a}}$} \\ \cline{2-8}
            & 600 \unit{\um}   & 455-515 & 60    & 2430$^{\textrm{c}}$ & 0.540$^{\textrm{c}}$ & 119.6$^{\textrm{c}}$ & 0.1196$^{\textrm{c}}$ \\ \cline{3-8}
            &+ 450 \unit{\um}{$^{\textrm{b}}$} & 650-710 & 60 & 2929$^{\textrm{c}}$ & 0.791$^{c}$ & 121.9$^{\textrm{c}}$ & 0.1219$^{\textrm{c}}$ \\ \hline
\end{tabular}
\caption{Estimated sensitivities for a typical track for different stages of the wSMA upgrade, under average weather (2 mm PWV) conditions (except as noted by ${\textrm{c}}$).\label{table:sens}}
\end{center}
\small{a. Assumes two bands operating simultaneously with only a single polarization per band.} \\
\small{b. 490 + 690 GHz receiver operating in the guest port position, in a separate cryostat.}\\
\small{c. Sensitivities estimated assuming the top quartile of weather, which roughly translates to 1 mm PWV.}
\end{table}

\subsection{The power of multi-band observing} A key capability for the wSMA will be simultaneous multi-band observing, which would be of tremendous value for both line and continuum observations across a broad range of scientific topics. Estimated sensitivities are shown in Table~\ref{table:sens}, which are derived assuming 8 antennas operating in median weather conditions (PWV$\approx$2 mm) unless otherwise noted, for a source at declination $\delta=0^{\circ}$, observed horizon-to-horizon down to an elevation limit of 20 degrees, with an assumed 67\% duty cycle for a total of 6.1 hours on source. A multi-band capable SMA would be an unique and valuable asset for astronomy, enabling the array to efficiently survey broadband emission, constrain the spectral energy distribution (SED) of continuum sources, and probe multiple line species of a number of molecules simultaneously, among other novel scientific goals and outputs. Moreover, the broad spectral coverage will further enable the SMA to take advantage of multi-frequency synthesis \cite{Sault1999} for continuum observations, which will radically improve the imaging capabilities of the array.

\section{OTHER UPGRADE OPPORTUNITIES}
\label{sec:other}
\subsection{Daytime and (near-)solar observing} One of the strengths of the Maunakea site is consistent weather suitable for submillimeter astronomy, which in terms of atmospheric opacity shows very little seasonal variation and only mild diurnal variation. As such, it is an exceptional site for supporting round-the-clock observations for sources of scientific interest. Studies conducted by SMA staff have demonstrated early success with correcting daytime phase instabilities, and with the addition of multi-band capabilities on the array, techniques such as frequency-phase transfer (FPT) would significantly expand the number of hours available for science on the array. 

The SMA antennas were originally intended to allow solar observing and have reflector surface finishes that scatter sunlight, greatly reducing the heat load from focused solar radiation.  Nevertheless, some of the as-built components near the focal region are susceptible to overheating, though readily replaced with more robust alternatives. Thus, a modest retrofit of the antenna structure would enable the array to engage in Solar observations, as well as objects much close to the Sun in the sky, such as comets and planetary bodies\cite{DeLuca2022}. The SMA's location on Hawai'i makes it particularly attractive for simultaneous millimeter and IR/optical solar observing with the Daniel K. Inouye Solar Telescope on Maui.

\subsection{Wide-field and on-the-fly mapping} With the growth of full-sky surveys over the next decade, the ability to support mapping of large contiguous areas, as well as measuring multiple distinct objects over wide fields, will remain a critical component for high-resolution science programs at submillimeter wavelengths. With a field of view of order 1 arcmin$^2$, fast slew speeds, and flexible observing software infrastructure, the SMA is an excellent tool for supporting large-area mapping, with surveys of up to 100 arcmin$^2$ easily enabled through mosaicked observations. Further improving these capabilities is a key component of future SMA science, and will be pursued both through observational planning, e.g. by enabling users to more easily script observations containing large mapping areas or multiple targets; and new observing modes, including the ability to conduct highly efficient “on-the-fly” (OTF) mapping of large areas, and to utilize single-dish data from the individual telescopes for mapping large scale structures. 

\subsection{Focal-plane arrays} In addition to the possibility of adding high frequency receivers to the SMA in the guest instrument position, there is also interest in deploying focal-plane array receiver there, to increase the widefield mapping speed of the SMA for specific projects. Although the fixed aperture sizes of the SMA's beam waveguide limit the options at longer wavelengths, initial optical designs have been carried out to demonstrate that up to 7 pixels could be employed at \qty{850}{\um}, and potentially more at shorter wavelengths. Used in combination with OTF mapping, high spatial and spectral resolution maps could be be made significantly faster than with single pixel receivers.

\subsection{Flux calibration accuracy} Flux calibration accuracy of 10-15\% is typical for submillimeter telescopes, making it a significant source of systematic error in many measurements, and potentially dominant on objects that are only moderately bright ($\gtrsim 1$ mJy). Thoughtful design and engineering, a wide field-of-view, modest size of the array (allowing for more careful calibration and characterization), and local astrophysical expertise, have allowed the SMA to routinely demonstrated flux accuracy significantly better than this typical level. 

Throughout the proposed upgrades, we will continue to incorporate design elements and techniques (e.g. the use of single-dish autocorrelation outputs from the correlator, and the inclusion of a cold calibration load in the wSMA receiver cryostat) to enhance the flux calibration accuracy of the SMA, with a goal of routinely acheiving 1\% flux calibration accuracy. Reliable flux calibration is  vital for characterizing short-lived or rapidly changing transient sources, where the frequency dependent brightness evolution of the source is key to understanding its nature. Such a capability would enable previously impossible investigations in several areas of astrophysics, including planetary science, star formation in nearby galaxies, and active galactic nuclei (AGN).

\acknowledgments % equivalent to \section*{ACKNOWLEDGMENTS}       
 
The Submillimeter Array is a joint project between the Smithsonian Astrophysical Observatory and the Academia Sinica Institute of Astronomy and Astrophysics and is funded by the Smithsonian Institution and the Academia Sinica. The SMA has received generous donations of FPGA chips for the SWARM correlator from AMD-Xilinx Inc., under the Xilinx University Program. Development of the VLBI features of SWARM were funded with SAO Internal Research \& Development funding, the NSF, and the Gordon and Betty Moore Foundation under GBMF3561. SMA EHT operations are funded by the NSF under MSI award 2034306. The SMA has benefited from technology shared under open-source license by the Collaboration for Astronomy Signal Processing and Electronics Research (CASPER) in building the SWARM correlator. This research has made use of NASA’s Astrophysics Data System.

We acknowledge the significance Maunakea has for the indigenous Hawaiian people; we are privileged to study the cosmos from its summit.

% References
\bibliography{main} % bibliography data in report.bib
\bibliographystyle{spiebib} % makes bibtex use spiebib.bst

\end{document}